\documentclass[%
 reprint,
 amsmath,amssymb,
 aps,
]{revtex4-2}
\usepackage{graphicx}
\usepackage{dcolumn}
\usepackage{bm}
\usepackage{graphicx}
\usepackage{dcolumn}
\usepackage{bm}
\usepackage{comment}

\usepackage{graphicx}
\usepackage{epsfig}
\usepackage{dcolumn}
\usepackage[up]{subfigure}
\usepackage{caption}
\usepackage{lipsum}
\usepackage{float}
\usepackage{dsfont}	
\usepackage{relsize}
\usepackage{verbatim}
\usepackage{blkarray, bigstrut}
\usepackage{xparse}

\newcommand\topstrut[1][1.2ex]{\setlength\bigstrutjot{#1}{\bigstrut[t]}}
\newcommand\botstrut[1][0.9ex]{\setlength\bigstrutjot{#1}{\bigstrut[b]}}
\usepackage{epstopdf}

\usepackage{amsfonts}

\usepackage{float}
\usepackage{ragged2e}
\usepackage{braket}
\usepackage{hyperref}
\usepackage{amsmath}
\usepackage{multirow}
\usepackage{mathtools}
\DeclarePairedDelimiter\ceil{\lceil}{\rceil}

\begin{document}
\newcommand{\Hsquare}{%
  \text{\fboxsep=-.2pt\fbox{\rule{0pt}{1ex}\rule{1ex}{0pt}}}%
}

\newcommand{\overl}[2]{\langle#1|#2\rangle}
\newcommand{\ph}{|\phi\rangle}
\newcommand{\ps}{|\psi\rangle}

\newcommand{\0}{|0\rangle}
\newcommand{\1}{|1\rangle}

\preprint{APS/123-QED}

\title{Moving Quantum States without SWAP via Intermediate Higher Dimensional Qudits}


\author{Amit Saha$^{1,2}$} 
\email{abamitsaha@gmail.com}
\author{Debasri Saha$^{1}$}
\author{Amlan Chakrabarti$^{1}$}
\email{asakc@caluniv.ac.in}

\affiliation{$^{1}$A. K. Choudhury School of Information Technology, University of Calcutta, India\\
$^{2}$ATOS, Pune, India}


\date{\today}
\begin{abstract}

Quantum algorithms can be realized in the form of a quantum circuit. To map quantum circuit for specific quantum algorithm to quantum hardware, qubit mapping is an imperative technique based on the qubit topology. Due to the neighbourhood constraint of qubit topology, the implementation of quantum algorithm rightly, is essential for moving information around in a quantum computer. Swapping of qubits using SWAP gate moves the quantum state between two qubits and solves the neighbourhood constraint of qubit topology. Though, one needs to decompose the SWAP gate into three CNOT gates to implement SWAP gate efficiently, but unwillingly quantum cost with respect to gate count and depth increases. In this paper, a new formalism of moving quantum states without using SWAP operation is introduced for the first time to the best of our knowledge. Moving quantum states through qubits have been attained with the adoption of temporary intermediate qudit states. This introduction of
intermediate qudit states has exhibited a three times reduction in quantum cost with respect to gate count and approximately two times reduction in respect to circuit depth compared to the state-of-the-art approach of SWAP gate insertion. Further, the proposed approach is generalized to any dimensional quantum system.
\end{abstract}

\maketitle


\section{\label{sec:level1}Introduction}

As it is experimentally quite established that the quantum computing system can be realized on various physical technologies, for example, continuous spin systems \cite{Bartlett_2002,Adcock_2016}, superconducting transmon technology \cite{PhysRevA.76.042319}, nuclear magnetic resonance \cite{Dogra_2014, Gedik_2015}, photonic systems \cite{Gao_2020},
 ion trap \cite{qutrit}, topological quantum systems \cite{Cui_2015first, Cui_2015} 
and molecular magnets \cite{Leuenberger_2001}, the physical implementation of quantum algorithms \cite{chuang} is now a blazing topic among the researchers for its asymptotic improvements \cite{Preskill_2018}. Transistors of classical computer deal with binary bits to accomplish information processing at the physical level. On the other hand, qubit technology is the base of quantum computers. A quantum system can have an infinite arity of discrete energy levels,  and hence, the fundamental physics behind the quantum system is not inherently binary. As per the real scenario, the impediment lies in the fact is that we need to control the system as per our requirements. The inclusion of additional discrete energy levels for the goal of computation enables us to realize the qudit technology quite predominantly, which makes the system more malleable with data storage and rapid processing of quantum information.

The first step towards implementation of a quantum algorithm is logic circuit synthesis. Since physical quantum computer only braces   
single-qubit gates and two-qubit gates \cite{barenco}, thus it becomes evident that the logical circuit synthesis must be decomposed into single-qubit gates and two-qubit gates so as to implement the algorithm on real quantum hardware devices. Every physical quantum computer has its own architectural design, qubit topology. A logic circuit design using only one qubit gates and two qubit gates does not suffice to be implemented physically. For this reason, there is the qubit mapping or qubit placement algorithm \cite{8342181, li2019tackling, tan2020optimal} based on qubit topology, which makes the implementation on physical quantum devices a reality. The operation involving two qubit gates are of most concern rather than single qubit gates while mapping them on physical devices, as the qubit topology may not support the placement of the required two physical qubits adjacently. To solve this constraint, ideally, SWAP gates \cite{PhysRevA.72.024303, PhysRevLett.102.020503} are used to move quantum states between two logical qubits. The idea is to exchange the qubits with repeated SWAP operations so that two logical qubits associated with two qubit gates can arrive at two adjacent physical qubits, but some additional cost is incurred.

In this paper, we have aspired to reduce the additional quantum cost that is incurred for the SWAP insertion \cite{wilmott2011generalized, wilmott2011generalized}. We propose a qubit-qudit approach \cite{PhysRevA.75.022313, lanyon, d0b63e4e43f844e9afe3288d494ae019, PhysRevA.88.034303} to move the quantum states through qubits to circumvent the SWAP operation.  This is a novel approach and achieves optimized gate cost and depth. One can simply have a higher dimensional quantum state for temporary use by easily introducing a discrete energy level. However, these higher dimensional quantum states are only present as intermediate states in a qudit system, while the input and output states are still remain qubits. We introduce the $\ket{2}$ and $\ket{3}$ quantum states as temporary storage of quaquad quantum system without hindering the fundamental operation of initialization and measurement on physical devices, since we are considering qubit system where two quantum states have to be temporarily stored. To the best of our knowledge, it is a first of its kind approach of moving quantum states through qubit without SWAP insertion via intermediate qudits, which is later extended to $d$-dimensional quantum system  with the use of $\ket{d}$, $\ket{d+1}$, $\dots$, $\ket{2d-1}$ quantum states as temporary storage. Our major contributions are the following:
\begin{itemize}
    \item With the use of temporary intermediate qudit states, moving quantum states through qubits have been studied for the first time to the best of our knowledge.

    \item These temporary intermediate qudit states help to reduce a significant number of quantum cost with respect to gate cost and circuit depth cost compared to the exchange of qubit states using SWAP gate.

    \item Further, we claim that with the help of temporary intermediate higher dimensional qudit states, quantum states can be moved through qudits in any  dimensional quantum system or $d$-ary quantum system with similar advancement with respect to quantum cost as binary quantum system, which makes our approach generalized in nature. 
\end{itemize}

The structure of the paper is as follows. Section 2 illustrates the SWAP gate and its usefulness. Section 3 exemplifies the methodology of moving quantum states through qubits using intermediate qudit with some example of circuit instances. Section 4 exhibits how the proposed method can be extended to any dimensional quantum system. Section 5 captures our conclusions with brief discussion.

\section{Background}
 The schematic diagram to represent quantum algorithm or quantum program is known as quantum circuit. Each line in the quantum circuit is denoted as a qubit and the operations i.e., quantum gates are represented by different blocks on the line \cite{Wang_2001, barenco}. There are mainly three basic cost metrics of a quantum circuit in quantum computing, viz. qubit cost, quantum gate cost/count (single or two qubit gates only) and depth of a circuit. The qubit cost and the quantum gate count are the number of qubits and the quantum gates respectively that are presented in a circuit. In a circuit, the path length for every case is an integer, which represent the number of gates to be executed in that path. Longest path in a circuit is the depth of the circuit. Since, all the qubits are not physically connected, the placement of the logical qubits needs to be rearranged to make them executable on the physical quantum devices. Fortunately, this can be done quite comprehensibly by using SWAP gates. Let's discuss more about SWAP gate.

\textbf{SWAP Gate:} In \cite{Garcia_Escartin_2013}, the authors have presented a SWAP gate implementation that interchanges the quantum states between two qubits. Let there be, the quantum states of two qubits as $\ph$ and $\ps$, then the SWAP gate will work as:
\begin{equation}
\mbox{SWAP}\ph\ps=\ps\ph.
\end{equation}

\begin{figure}[ht!]
\centering
\includegraphics[scale=.8]{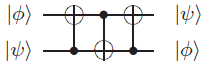}
\caption{CNOT swapping circuit.\label{CNOTSWAP}} 
\end{figure}

Three CNOT gates constitute the SWAP operation as in Figure \ref{CNOTSWAP}. CNOT is the two-qubit universal gate that has a control qubit, portrayed as a black dot ($\bullet$), and a target qubit, portrayed with the XOR symbol ($\oplus$). If the control qubit of a CNOT gate is in quantum state $\1$, the target qubit's value alters from $\0$ to $\1$ or/and from $\1$ to $\0$. The CNOT gate can be mathematically illustrated as 
\begin{equation}
\mbox{CNOT}\ket{x}\ket{y}=\ket{x}\ket{x\oplus y}. 
\end{equation}
The XOR operation is modulo 2 addition where target qubit is incremented by $1 \ (\text{mod } 2)$ only when the control qubit value is $1$.

In following steps (Equation 3, 4, 5), how the concatenation of three CNOT gates implements the SWAP operation is shown: 
\begin{equation}
\ket{x}\ket{y}\stackrel{\mbox{\scriptsize{CNOT}}_{q_1,q_0}}{\longrightarrow}\ket{x\oplus y}\ket{y},
\end{equation}
\begin{equation}
\ket{x\oplus y}\ket{y}\stackrel{\mbox{\scriptsize{CNOT}}_{q_0,q_1}}{\longrightarrow}\ket{x\oplus y}\ket{y\oplus x \oplus y}=\ket{x\oplus y}\ket{x },
\end{equation}
\begin{equation}
\ket{x\oplus y}\ket{x}\stackrel{\mbox{\scriptsize{CNOT}}_{q_1,q_0}}{\longrightarrow}\ket{x\oplus y \oplus x}\ket{x}=\ket{y}\ket{x}. 
\end{equation}
Here, by convention, $\mbox{CNOT}_{i,j}$ is a CNOT gate controlled by qubit $i$ and with qubit $j$ as target. In this paper, qubits are labeled serially as \{$q_0$, $q_1$, $\dots$, $q_n$\}. The unitary transformation of the quantum states that we have illustrated here must give desired result even if the quantum states are superposed.

The alternative construction of three concatenated CNOT gates to implement SWAP operation is shown below: 

\begin{figure}[ht!]
\centering
\includegraphics[scale=.8]{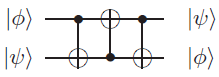}
\caption{Alternative configuration for the CNOT swapping circuit.\label{altCNOTSWAP}} 
\end{figure}

\textbf{Qubit Mapping Problem:} For explaining this qubit mapping
problem with the help of SWAP insertion, we have considered an example as shown in Figure 3(a). In Figure 3(b), a 3-qubit topology is used as the hardware
platform. Two-qubit gates are executable
on the following adjacent physical qubits: \{${P_0, P_1}$\}, \{${P_1, P_2}$\} and not on \{${P_0, P_2}$\}.
Now, suppose we have a CNOT to be
executed on this 3-qubit device. This quantum circuit consists
of several other gates (as shown in Figure 3(a) as block of gates). Assuming the
initial logical-to-physical qubits mapping is \{${q_0 \longrightarrow P_0, q_1 \longrightarrow P_1, q_2 \longrightarrow P_2}$\}. We can find that CNOT gate ($g_r$) as shown in Figure 3(a) cannot
be executed because the corresponding qubit pairs are not
connected on the device. We need to change the qubit mapping during execution,
and make the CNOT gate executable.

\begin{figure}[ht!]
\centering
\includegraphics[scale=0.4]{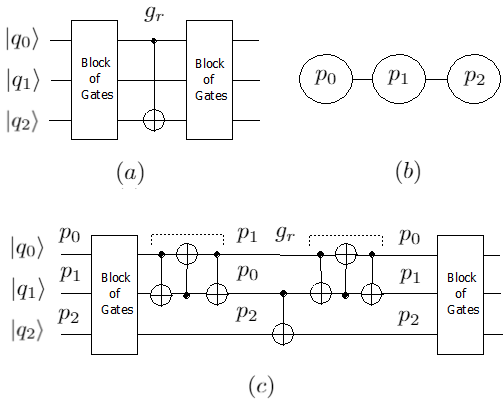}
\caption{$(a)$ Example: circuit $(b)$ Example: qubit topology $(c)$ SWAP insertion}
\label{exampleswapinsertion}
\end{figure}

To overcome this issue, we
employ SWAP operation to change the qubit mapping by
exchanging the states between two qubits. It consists of three
CNOT gates as shown in Figure 1 and Figure 2. Figure 3(c) shows that the
updated quantum circuit is now executable after we insert
one SWAP operation between $q_0$ and $q_1$ as shown dotted. After the inserted SWAP, mapping is updated to
\{${q_0 \longrightarrow P_1, q_1 \longrightarrow P_0, q_2 \longrightarrow P_2}$\}. Now, the CNOT gate can be executed under this updated
mapping. For further execution of remaining block of gates, we need to again apply SWAP operation between $q_0$ and $q_1$ as shown dotted in Figure 3(c) to get back to the previous logical-to-physical qubits mapping, which is \{${q_0 \longrightarrow P_0, q_1 \longrightarrow P_1, q_2 \longrightarrow P_2}$\}.

With the introduction of additional SWAPs in the quantum circuit,
 all the two-qubit gate dependencies can be solved and 
a hardware-compliant circuit with unchanged original
functionality is generated. Furthermore, 
insertion of SWAPs in the quantum circuit will lead to several
problems, because of the limitations of quantum devices. There is an increase in the  number of operations in the circuit. The overall error rate increases as the operations are not perfect and noise will also be introduced. There might also be increase in depth of the circuit, i.e. there will be an increase in the total execution time and due to qubit decoherence, there will be an accumulation of 
too much error. If we compare the original circuit and the updated circuit in
Figure 3(a) and Figure 3(c), the number of gates increases from 1
to 7 and the circuit depth is also increased from 1 to 7.  Significant overhead in terms of fidelity and
execution time will be brought with additional SWAPs. Thus, in order to reduce the overall error rate as well as total execution time for the final hardware-complaint circuit, we look forward to discard the additional SWAPs.

\section{Moving qubit states via Intermediate Qudits}

The most important aspect of our proposed work is the moving quantum states through qubits without SWAP so that overall error rate can be optimized. In this regard,  the states $\ket{2}$ and $\ket{3}$ of higher dimensional quaquad system have been used in the intermediate levels during the computation. Since we keep  input/output as binary, it enables these circuit constructions to act similarly as any already existing binary qubit-only circuits. Figure 4 describes how this can be achieved for the circuit shown in Figure 3(a) based on qubit topology as shown in Figure 3(b).

\begin{figure}[ht!]
\centering
\includegraphics[scale=0.55]{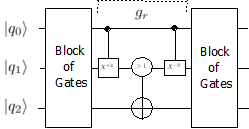}
\caption{Moving of quantum state via intermediate qudits}
\label{exampleintermediatequbits}
\end{figure}

Although physical systems in classical hardware are typically binary, but, common quantum hardware, such as in superconducting and trapped ion computers, has an infinite spectrum of discrete
energy levels \cite{gokhale2019asymptotic}. Quantum hardware may be configured to manipulate the lowest four energy levels by operating on quaquads. In general, such a computer could be configured to operate on any number of $d$ levels.  Qudit gates have already been successfully implemented \cite{PhysRevA.76.042319, qutrit} indicating it is possible to consider higher level systems apart from qubit only systems. Thus, the question of higher states beyond the standard two, being implemented and performed no longer stands strong. Since, conventional binary quantum gates \cite{barenco} are not capable enough to get access to higher dimensional quantum states, hence new qubit-qudit quantum gates are needed to be introduced. First, let us consider an increment gate as $C^{+2}_{X}$ ($C$ : Control; $X$ : NOT), where $+2$ denotes that the target qubit is incremented by $2 \ (\text{mod } 4)$ as 4-ary quantum system considered, only when the control qubit value is $1$. For visualization of the $C^{+2}_{X}$ gate, we have used a 'Black dot' ($\bullet$) to represent the control, and a 'rectangle' ($\Hsquare$) to represent the target. '$X^{+2}$' in the target box represents the increment operator. The mathematical representation of the $C^{+2}_{X}$ gate is as follows:

\begin{equation}
\resizebox{.75\hsize}{!}{$ C^{+2}_{X}\ket{x}\ket{y}= \left\{\begin{array}{ll}\mbox{$\ket{x}\ket{(y+2) \% 4},$} & \mbox{if $x = 1$};\\
\mbox{$\ket{x}\ket{y}$,} & \mbox{otherwise}.\\
\end{array}\right. $}
\end{equation}

Since we are working with qubit-qudit approach, this does not encapsulate the complete scenario, so we need to describe it with a
matrix instead. The $(8 \times 8)$ unitary matrix representation of the $C^{+2}_{X}$ gate is as follows:

\begin{equation*}
C^{+2}_{X} =
\begin{blockarray}{l c c c c c c c c}
& 00 & 01 & 02 & 03 & 10 & 11 & 12 & 13\\[-0.8ex] 
 \begin{block}{l [c c c c c c c c]}
 00 & 1 & 0 & 0 & 0 & 0 & 0 & 0 & 0 \topstrut\\
01 &   0 & 1 & 0 & 0 & 0 & 0 & 0 & 0 \\
02 &   0 & 0 & 1 & 0 & 0 & 0 & 0 & 0  \\
03 &   0 & 0 & 0 & 1 & 0 & 0 & 0 & 0  \\
10 &   0 & 0 & 0 & 0 & 0 & 0 & 1 & 0  \\
11 &   0 & 0 & 0 & 0 & 0 & 0 & 0 & 1  \\
12 &   0 & 0 & 0 & 0 & 1 & 0 & 0 & 0  \\
13 &   0 & 0 & 0 & 0 & 0 & 1 & 0 & 0 \botstrut \\
\end{block}
\end{blockarray}
 \end{equation*}

  This gate operation is performed on the first and the second qubit as shown in Figure 4, where first qubit is the control and the second qubit is the target. This upgrades the second qubit to $\ket{2}$ or $\ket{3}$ by availing the higher dimensional Hilbert space as temporary storage if and only if the first qubit was $\ket{1}$.
  
  Then, a conditional CNOT gate as $C^{+1}_{X_c}$, where $c$ : conditional operator; and $+1$ denotes that the target qubit is incremented by $1 \ (\text{mod } 2)$ as target qubit is in binary quantum system, if and only if the control qubit value is greater than $1$, is applied to the target qubit i.e., third qubit and the second qubit as control. In the schematic design of $C^{+1}_{X_c}$ gate, we have used '$>1$' in the conditional control circle (O) to represent the qubit control, and XOR ($\oplus$) in the target qubit to represent the conditional CNOT operator. The mathematical representation of the $C^{+1}_{X_c}$ gate is as follows:

  \begin{equation}
\resizebox{.75\hsize}{!}{$ C^{+1}_{X_c}\ket{x}\ket{y}= \left\{\begin{array}{ll}\mbox{$\ket{x}\ket{(y+1) \% 2},$} & \mbox{if $x > 1$};\\
\mbox{$\ket{x}\ket{y}$,} & \mbox{otherwise}.\\
\end{array}\right. $}
\end{equation}

The $(8 \times 8)$ unitary matrix representation of the $C^{+1}_{X_c}$ gate is as follows:

\begin{equation*}
C^{+1}_{X_c} = \begin{blockarray}{l c c c c c c c c}
& 00 & 01 & 10 & 11 & 20 & 21 & 30 & 31\\[-0.8ex] 
 \begin{block}{l [c c c c c c c c]}
 00 & 1 & 0 & 0 & 0 & 0 & 0 & 0 & 0 \topstrut\\
01 &   0 & 1 & 0 & 0 & 0 & 0 & 0 & 0 \\
10 &   0 & 0 & 1 & 0 & 0 & 0 & 0 & 0  \\
11 &   0 & 0 & 0 & 1 & 0 & 0 & 0 & 0  \\
20 &   0 & 0 & 0 & 0 & 0 & 1 & 0 & 0  \\
21 &   0 & 0 & 0 & 0 & 1 & 0 & 0 & 0  \\
30 &   0 & 0 & 0 & 0 & 0 & 0 & 0 & 1  \\
31 &   0 & 0 & 0 & 0 & 0 & 0 & 1 & 0 \botstrut \\
\end{block}
\end{blockarray}
  \end{equation*}
 
The conditional CNOT gate is executed only when the second qubit were $\ket{2}$ or $\ket{3}$, as expected, it would happen only when the first qubit was $\ket{1}$ state.  The controls are reinstated to their original states by a $C^{-2}_{X}$ gate i.e., inverse of $C^{+2}_{X}$ gate, which reverses the effect of the first gate. Thus the $\ket{2}$ and $\ket{3}$ state from $4$-ary quantum system can be used instead of SWAP to store temporary information, which is the most important aspect in this circuit composition. As shown in Figure 4, the input of the circuit can be a form of $\sum_{x,y,t = 0}^{1}\alpha_{x, y, t}\ket{x}\ket{y}\ket{t}$
where the first two are qubits $q_0$ and $q_1$, and the target qubit $t$ is $q_2$,  and $\alpha_{x, y, t} \in \mathbb{C}$ and $\sum_{x,y,t =0}^{1} |\alpha_{x, y, t}|^2 = 1$. Here, we show the action of the proposed gates (Equation 8, 9, 10, 11) on such a superposition by ignoring the 'block of gates' for the sake of simplicity and understanding now and then.
\begin{widetext}
\begin{eqnarray}
& & \sum_{x,y,t = 0}^{1}\alpha_{x, y, t}\ket{x}\ket{y}\ket{t}\\
&\stackrel{\mbox{\scriptsize{$C^{+2}_{X}$}}_{q_0,q_1}}{\longrightarrow}& \sum_{x=0,y,t}\alpha_{x=0,y,t}\ket{x}\ket{y}\ket{t} + \sum_{x=1,y,t}\alpha_{x=1,y,t}\ket{1}\ket{(y+2)\%4}\ket{t}\\
&\stackrel{\mbox{\scriptsize{$C^{+1}_{X_c}$}}_{q_1,q_2}}{\longrightarrow}& \sum_{x=0,y,t}\alpha_{x=0,y,t}\ket{x}\ket{y}\ket{t} + \sum_{x=1,y,t}\alpha_{x=1,y,t}\ket{1}\ket{(y+2)\%4}\ket{(t+1)\%2}\\
&\stackrel{\mbox{\scriptsize{$C^{-2}_{X}$}}_{q_0,q_1}}{\longrightarrow}& \sum_{x=0,y,t}\alpha_{x=0,y,t}\ket{x}\ket{y}\ket{t} + \sum_{x=1,y,t}\alpha_{x=1,y,t}\ket{1}\ket{(y}\ket{(t+1)\%2}
\end{eqnarray}
\end{widetext}

One more example of small 4-qubit size (Figure 5(a)) is used for explaining the changes of circuit realization if there is an increase in the number of qubit. In Figure 5(b), a 4-qubit device model is utilized as the hardware
platform. Two-qubit gates are executable on the following physical qubit pairs: \{${P_0, P_1}$\}, \{${P_1, P_2}$\} and \{${P_2, P_3}$\} not on \{${P_0, P_2}$\}, \{${P_0, P_3}$\} and \{${P_1, P_3}$\}. Let there be a CNOT gate controlled  by  qubit $q_0$ and qubit $q_3$ as target to be executed on this 4-qubit device.  Assuming the initial logical-to-physical qubits mapping is \{${q_0 \longrightarrow P_0, q_1 \longrightarrow P_1, q_2 \longrightarrow P_2}$\}, the CNOT gate ($g_r$) as in Figure 5(a) cannot be executed due to the corresponding qubit pairs being disconnected on the device. Hence, the qubit mapping needs to be changed  during execution and the CNOT gate must be made executable.

\begin{figure}[!h]
\centering
\includegraphics[scale=0.4]{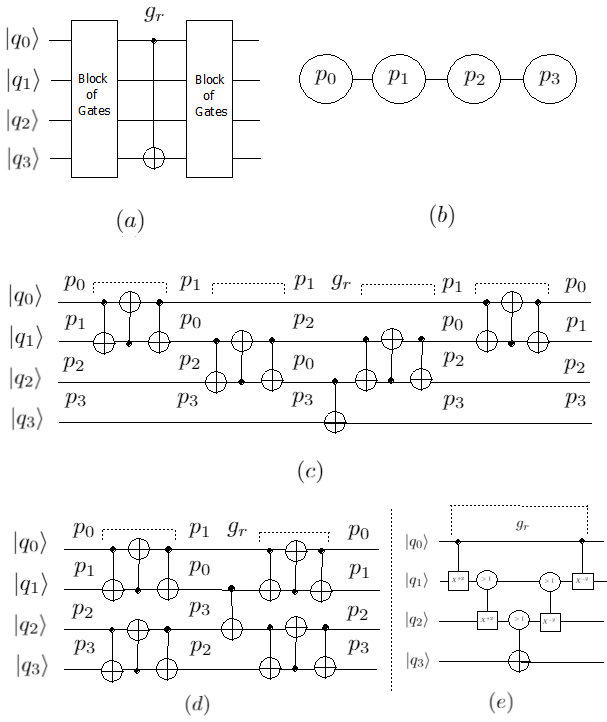}
\caption{$(a)$ Example: circuit; $(b)$ Example: qubit topology; $(c)$ SWAP insertion; $(d)$ SWAP Depth optimization $(e)$ Proposed approach}
\label{exampleswapinsertion}
\end{figure}

Conventionally, SWAP operation is needed to be employed for changing the qubit mapping with
exchange of the quantum states between two qubits as in figure 3(c). Likewise, Figure 5(c) shows that the
updated quantum circuit is now executable after we insert
two SWAP operations between '$q_0$ and $q_1$' and  '$q_1$ and $q_2$' as shown dotted in Figure 5(c). After the inserted SWAPs, logical qubit to physical qubit mapping is updated to
\{${q_0 \longrightarrow P_1, q_1 \longrightarrow P_2, q_2 \longrightarrow P_0, q_3 \longrightarrow P_3}$\}. Now, the CNOT gate can be executed following this updated mapping. For further execution of remaining block of gates, we need to again apply SWAP operation between '$q_0$ and $q_1$' and '$q_2$ and $q_3$' as shown dotted in Figure 5(c) to get back to the previous logical-to-physical qubits mapping. As per Gidney's work \cite{gidney}, the depth of circuit shown in 5(c) can be optimized to a circuit shown in 5(d). This optimization shows that two series of SWAP gates can be implemented parallely. Two parallely inserted SWAPs update the mapping to \{${q_0 \longrightarrow P_1, q_1 \longrightarrow P_0, q_2 \longrightarrow P_3, q_3 \longrightarrow P_2}$\}. Now, the CNOT gate ($g_r$) can be executed under this updated mapping. Similarly, miror circuit can be applied parallely as well, which improves the depth of a circuit but the gate cost remains unchanged to the previous approach of circuit realization as shown in 5(c).

To eradicate the SWAP operation from 5(d), we need to introduce another new gate that is $C^{+2}_{X_c}$ as the number of qubits have increased from three to four. $C^{+2}_{X_c}$ gate along with previously proposed $C^{+2}_{X}$ and $C^{+1}_{X_c}$ gates will lead to execute the circuit shown in Figure 5(a) based on the qubit topology as shown in Figure 5(b). A conditional increment gate as $C^{+2}_{X_c}$, where $+2$ denotes that the target qubit is incremented by $2 \ (\text{mod } 4)$ as 4-ary quantum system considered, only when the control qubit value is greater than $1$. In the design of the gate, '$>1$'  has been used in the conditional control circle (O) to represent the qubit control, and '$X^{+2}$' in the target rectangular box ($\Hsquare$) to represent the increment operator. The mathematical representation of the $C^{+2}_{X_c}$ gate is as follows:

  \begin{equation}
\resizebox{.75\hsize}{!}{$ C^{+2}_{X_c}\ket{x}\ket{y}= \left\{\begin{array}{ll}\mbox{$\ket{x}\ket{(y+2) \% 4},$} & \mbox{if $x > 1$};\\
\mbox{$\ket{x}\ket{y}$,} & \mbox{otherwise}.\\
\end{array}\right. $}
\end{equation}

The $(16 \times 16)$ unitary matrix representation of the $C^{+2}_{X_c}$ gate is as follows:


\begin{equation*}
C^{+2}_{X_c} = \begin{blockarray}{l c c c c c c c c c c c c c c c c}
& 00 & 01 & 02 & 03 & 10 & 11 & 12 & 13 & 20 & 21 & 22 & 23 & 30 & 31 & 32 & 33\\[-0.8ex]
 \begin{block}{l [c c c c c c c c c c c c c c c c]}
00 & 1 & 0 & 0 & 0 & 0 & 0 & 0 & 0 & 0 & 0 & 0 & 0 & 0 & 0 & 0 & 0 \topstrut\\
01 & 0 & 1 & 0 & 0 & 0 & 0 & 0 & 0 & 0 & 0 & 0 & 0 & 0 & 0 & 0 & 0\\
02 & 0 & 0 & 1 & 0 & 0 & 0 & 0 & 0 & 0 & 0 & 0 & 0 & 0 & 0 & 0 & 0\\
03 & 0 & 0 & 0 & 1 & 0 & 0 & 0 & 0 & 0 & 0 & 0 & 0 & 0 & 0 & 0 & 0\\
10 & 0 & 0 & 0 & 0 & 1 & 0 & 0 & 0 & 0 & 0 & 0 & 0 & 0 & 0 & 0 & 0\\
11 & 0 & 0 & 0 & 0 & 0 & 1 & 0 & 0 & 0 & 0 & 0 & 0 & 0 & 0 & 0 & 0\\
12 & 0 & 0 & 0 & 0 & 0 & 0 & 1 & 0 & 0 & 0 & 0 & 0 & 0 & 0 & 0 & 0\\
13 & 0 & 0 & 0 & 0 & 0 & 0 & 0 & 1 & 0 & 0 & 0 & 0 & 0 & 0 & 0 & 0\\
20 & 0 & 0 & 0 & 0 & 0 & 0 & 0 & 0 & 0 & 0 & 1 & 0 & 0 & 0 & 0 & 0\\
21 & 0 & 0 & 0 & 0 & 0 & 0 & 0 & 0 & 0 & 0 & 0 & 1 & 0 & 0 & 0 & 0\\
22 & 0 & 0 & 0 & 0 & 0 & 0 & 0 & 0 & 1 & 0 & 0 & 0 & 0 & 0 & 0 & 0\\
23 & 0 & 0 & 0 & 0 & 0 & 0 & 0 & 0 & 0 & 1 & 0 & 0 & 0 & 0 & 0 & 0\\
30 & 0 & 0 & 0 & 0 & 0 & 0 & 0 & 0 & 0 & 0 & 0 & 0 & 0 & 0 & 1 & 0\\
31 & 0 & 0 & 0 & 0 & 0 & 0 & 0 & 0 & 0 & 0 & 0 & 0 & 0 & 0 & 0 & 1\\
32 & 0 & 0 & 0 & 0 & 0 & 0 & 0 & 0 & 0 & 0 & 0 & 0 & 1 & 0 & 0 & 0\\
33 & 0 & 0 & 0 & 0 & 0 & 0 & 0 & 0 & 0 & 0 & 0 & 0 & 0 & 1 & 0 & 0 \botstrut\\
\end{block}
\end{blockarray}
  \end{equation*}


Figure 5(e) shows how the proposed gates can execute the circuit shown in Figure 5(a) based on qubit topology as shown in Figure 5(b) via temporary intermediate qudits. The initialization of Figure 5(e) can be expressed as 
$\sum_{x,y,z,t = 0}^{1}\alpha_{x, y, z, t}\ket{x}\ket{y}\ket{z}\ket{t}$, where the first three are qubits $q_0$, $q_1$ and $q_2$ and the target qubit $t$ is $q_3$,  and $\alpha_{x, y, z, t} \in \mathbb{C}$ and $\sum_{x,y,z,t =0}^{1} |\alpha_{x, y, z, t}|^2 = 1$. Here, we show the action of a proposed gates on such a superposition. At first, the $C^{+2}_{X}$ gate operation is performed on the first and the second qubit as illustrated in Equation 14, where first qubit is the control and the second qubit is the target. This upgrades the second qubit to $\ket{2}$ or $\ket{3}$ by availing the higher dimensional Hilbert space as temporary storage if and only if the first qubit was $\ket{1}$. Next, the $C^{+2}_{X_c}$ gate operation is performed on the second and the third qubit as illustrated in Equation 15, where second qubit is the control and the third qubit is the target. This upgrades the third qubit to $\ket{2}$ or $\ket{3}$ by availing the higher dimensional Hilbert space as temporary storage if and only if the second qubit was $\ket{2}$ or $\ket{3}$. Finally, a conditional CNOT $C^{+1}_{X_c}$ is applied to the target qubit $q_3$ and the third qubit as control as describes in Equation 16. This gate will be executed only when the third qubit were $\ket{2}$ or $\ket{3}$, as expected and as discussed earlier, it would happen only when the first qubit was $\ket{1}$ state.  The controls are reinstated to their original states by applying $C^{-2}_{X_c}$ gate followed by $C^{-2}_{X}$ gate, which reverses the effect of the first and second gate. Thus the $\ket{2}$ and $\ket{3}$ state from $4$-ary quantum system can be used instead of SWAP to store temporary information in $q_1$ and $q_2$ qubits, which is the most important aspect in this 4-qubit circuit composition.
\begin{widetext}
\begin{eqnarray}
& & \sum_{x,y,z,t}^{1}\alpha_{x, y, z t}\ket{x}\ket{y}\ket{z}\ket{t}\\
&\stackrel{\mbox{\scriptsize{$C^{+2}_{X}$}}_{q_0,q_1}}{\longrightarrow}& \sum_{x=0,y,z,t}\alpha_{x=0,y,z,t}\ket{x}\ket{y}\ket{z}\ket{t} + \sum_{x=1,y,z,t}\alpha_{x=1,y,z,t}\ket{1}\ket{(y+2)\%4}\ket{z}\ket{t}\\
&\stackrel{\mbox{\scriptsize{$C^{+2}_{X_c}$}}_{q_1,q_2}}{\longrightarrow}& \sum_{x=0,y,z,t}\alpha_{x=0,y,z,t}\ket{x}\ket{y}\ket{z}\ket{t} + \sum_{x=1,y,z,t}\alpha_{x=1,y,z}\ket{1}\ket{(y+2)\%4}\ket{(z+2)\%4}\ket{t}\\
&\stackrel{\mbox{\scriptsize{$C^{+1}_{X_c}$}}_{q_2,q_3}}{\longrightarrow}& \sum_{x=0,y,z,t}\alpha_{x=0,y,z,t}\ket{x}\ket{y}\ket{z}\ket{t} + \sum_{x=1,y,z,t}\alpha_{x=1,y,z,t}\ket{1}\ket{(y+2)\%4}\ket{(z+2)\%4}\ket{(t+1)\%2}\\
&\stackrel{\mbox{\scriptsize{$C^{-2}_{X_c}$}}_{q_1,q_2}}{\longrightarrow}& \sum_{x=0,y,z,t}\alpha_{x=0,y,z}\ket{x}\ket{y}\ket{z}\ket{t} + \sum_{x=1,y,z,t}\alpha_{x=1,y,z}\ket{1}\ket{(y+2)\%4}\ket{z}\ket{(t+1)\%2}\\
&\stackrel{\mbox{\scriptsize{$C^{-2}_{X}$}}_{q_0,q_1}}{\longrightarrow}& \sum_{x=0,y,z,t}\alpha_{x=0,y,z,t}\ket{x}\ket{y}\ket{z}\ket{t} + \sum_{x=1,y,z,t}\alpha_{x=1,y,z}\ket{1}\ket{(y}\ket{z}\ket{(t+1)\%2}
\end{eqnarray}
\end{widetext}

In Figure 6(a) and Figure 6(b), we have shown examples of a CNOT gate with 5-qubits and 6-qubits respectively. In these examples as in earlier example, let's assume  CNOT's control is in the first qubit and target is in the last qubit. For executing the CNOT gate, we need to move qubit states through the intermediate qubits by accessing the higher dimensional Hilbert space. In Figure 6(a) and 6(b), we show that these three proposed gates $C^{+2}_{X}$, $C^{+2}_{X_c}$ and $C^{+1}_{X_c}$ are sufficient to execute a CNOT gate in higher qubit system as well. From this background, it can be inferred, for any higher $n$-qubit system ($q_1$, $q_2$, $q_3$, $\dots$, $q_{n-1}$, $q_n$, where the two-qubit gate is involved between $q_1$ and $q_n$), the proposed three gates can be used for moving quantum states for physical implementation without the use of SWAP gate. We can conclude that, $C^{+2}_{X}$ gate is used between the first and second qubits i.e., $q_1$ and $q_2$, for intermediate operations $C^{+2}_{X_c}$ gate is used on \{($q_2$, $q_3$), ($q_3$, $q_4$), $\dots$, ($q_{n-2}$, $q_{n-1}$)\} and finally $C^{+1}_{X_c}$ is executed with the control qubit $q_{n-1}$ and target qubit $q_n$.

\begin{figure}[ht!]
\centering
\includegraphics[scale=0.65]{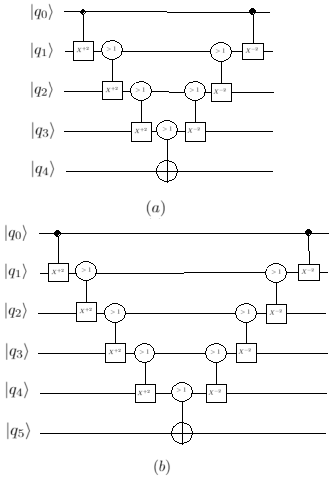}
\caption{$(a)$ 5-qubit circuit $(b)$ 6-qubit circuit}
\label{exampleswapinsertion}
\end{figure}

\section{Moving $d$-dimensional quantum states via higher dimensional qudits}

In this section, we consider the implementation of our proposed qubit-qudit method generalised to any dimension as qudit-higher dimensional qudit method. Qudit technology is concerned with $d$-ary quantum systems, where $d > 2$ \cite{Muthukrishnan_2000, Di_2013}. We graduate to qudits for providing a larger state space and simultaneous multiple control operations,  which in the long run reduce the circuit complexity and uplift the efficiency of quantum algorithms \cite{qft, groverd, 9410395}. For example, $N$ qubits can be depicted as $\frac{N}{log_2 {d}}$ qudits, which straightway reduces a $log_2 {d}$-factor from the run-time of a quantum algorithm \cite{Wang_2020}. An akin construction of proposed binary gates using qudit have been extended for $d$-ary quantum system by generalising the $C^{+2}_{X_c}$, $C^{+2}_{X}$ and $C^{+1}_{X_c}$ gates. The aim is to move the $d$-dimensional quantum states through qudits by accessing the higher dimensional quantum space as temporary storage. As we have shown binary quantum system needs to access $\ket{2}$ and $\ket{3}$ of quaquad system, likewise, it generalizes for $d$-dimensional quantum system by accessing additional $d$-dimensional Hilbert space, since $d$ quantum states have to be temporarily stored. with the use of $\ket{d}$, $\ket{d+1}$, $\dots$, $\ket{2d-1}$ quantum states of $2d$-dimensional quantum system as temporary storage, we can avoid SWAP gate in qudit system to get a solution of our objective. Before discussing more about our proposed method, let's enlighten about the SWAP gate in qudit system \cite{bala, wilmott2011swapping, Wilmott_2014}.

\textbf{SWAP Gate in Qudit System:} In \cite{Garcia_Escartin_2013}, the author proposed a gate $C\!\widetilde{X}$, a generalization of CNOT gate in qudit system, which generally acts on qudits $\ket{x}$ and $\ket{y}$ from the basis $\left\{ \ket{0}, \ket{1}, \ldots, \ket{d-1}\right\}$ so that
\begin{equation}
C\!\widetilde{X}\ket{x}\ket{y}=\ket{x}\ket{-x-y}.
\end{equation}
$\ket{-x-y}$ denotes a state $\ket{i}$ in the range $i=0,\ldots,d-1$ with $i=-x-y \mod d$.

\begin{figure}[ht!]
\centering
\includegraphics[scale=.8]{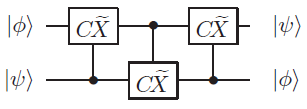}
\caption{Qudit swapping circuit.\label{CXSWAP}} 
\end{figure}

\begin{figure}[ht!]
\centering
\includegraphics[scale=.8]{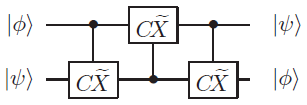}
\caption{Alternative configuration for the qudit swapping circuit.\label{altCXSWAP}} 
\end{figure}

The SWAP gate shown in Figure \ref{CNOTSWAP} has been extended to qudit system using three $C\!\widetilde{X}$ gates, which is described in Figure \ref{CXSWAP}. If $C\!\widetilde{X}_{i,j}$ is a $C\!\widetilde{X}$ gate where the control is qudit $i$ and the target qudit $j$, the qudit SWAP gate that is evolved through
\begin{equation}
\ket{x}\ket{y}\stackrel{C\!\widetilde{X}_{q_2,q_1}}{\longrightarrow}\ket{-x-y}\ket{y},
\end{equation}
\begin{equation}
\ket{-x-y}\ket{y}\stackrel{C\!\widetilde{X}_{q_1,q_2}}{\longrightarrow}\ket{-x-y}\ket{x+y-y}=\ket{-x-y}\ket{x},
\end{equation}
\begin{equation}
\ket{-x-y}\ket{x}\stackrel{C\!\widetilde{X}_{q_2,q_1}}{\longrightarrow}\ket{-x+x+y}\ket{x}=\ket{y}\ket{x}.
\end{equation}
This SWAP operation for qudit system must be acted upon appropriately for any possible arbitrary superposed input qudit with quantum state from $\left\{ \ket{0}, \ket{1}, \ldots, \ket{d-1}\right\}$. Figure \ref{altCXSWAP} portrays an alternative circuit construction of SWAP gate in qudit system with the help of three concatenated $C\!\widetilde{X}$ gates.

As mentioned earlier, the proposed binary gates are needed to be generalized for d-ary quantum system. Let us consider a generalized increment gate for $d$-ary quantum system as $C^{+d}_{X}$, where $+d$ denotes that the target qudit is incremented by $d \ (\text{mod } 2d)$ as $2d$-ary quantum system considered, if and only if the control qudit value is $d-1$. For visualization of the $C^{+d}_{X}$ gate, we have used a 'Black dot' ($\bullet$) to represent the control, and a 'rectangle' ($\Hsquare$) to represent the target. '$X^{+d}$' in the target box represents the increment operator. The mathematical representation of the $C^{+d}_{X}$ gate is as follows:

\begin{equation}
\resizebox{.75\hsize}{!}{$ C^{+d}_{X}\ket{x}\ket{y}= \left\{\begin{array}{ll}\mbox{$\ket{x}\ket{(y+d) \% 2d},$} & \mbox{if $x =d-1$};\\
\mbox{$\ket{x}\ket{y}$,} & \mbox{otherwise}.\\
\end{array}\right. $}
\end{equation}

A conditional increment gate can be extended to a generalized conditional increment gate for $d$-ary quantum system as $C^{+d}_{X_c}$, where $+d$ denotes that the target qudit is incremented by $d \ (\text{mod } 2d)$ as $2d$-ary quantum system considered, only when the control qudit value is greater than $d-1$. In the design of the $C^{+d}_{X_c}$ gate, '$>d-1$'  has been used in the conditional control circle (O) to represent the qudit control, and '$X^{+d}_c$' in the target rectangular box ($\Hsquare$) to represent the increment operator. The mathematical representation of the $C^{+d}_{X_c}$ gate is as follows:

 \begin{equation}
\resizebox{.75\hsize}{!}{$ C^{+d}_{X_c}\ket{x}\ket{y}= \left\{\begin{array}{ll}\mbox{$\ket{x}\ket{(y+d) \% 2d},$} & \mbox{if $x > d-1$};\\
\mbox{$\ket{x}\ket{y}$,} & \mbox{otherwise}.\\
\end{array}\right. $}
\end{equation}

In similar way, a conditional CNOT gate can be extended to $d$-ary quantum system. The generalized conditional CNOT gate can be defined for $d$-ary quantum system as $C^{+a}_{X_c}$, where $+a$ denotes that the target qudit is incremented by $a \ (\text{mod } d)$ as $d$-ary quantum system considered, if and only if the control qudit value is greater than $d-1$ while $1\leq a \leq d-1$. In the schematic design of $C^{+a}_{X_c}$ gate, we have used '$>d-1$' in the conditional control circle (O) to represent the qudit control, and '$X^{+a}_c$' in the target rectangular box ($\Hsquare$) to represent the conditional CNOT operator. The mathematical representation of the $C^{+a}_{X_c}$ gate is as follows:

 \begin{equation}
\resizebox{.75\hsize}{!}{$ C^{+a}_{X_c}\ket{x}\ket{y}= \left\{\begin{array}{ll}\mbox{$\ket{x}\ket{(y+a) \% d},$} & \mbox{if $x > d-1$};\\
\mbox{$\ket{x}\ket{y}$,} & \mbox{otherwise}.\\
\end{array}\right. $}
\end{equation}

\begin{figure}[]
\centering
\includegraphics[scale=0.4]{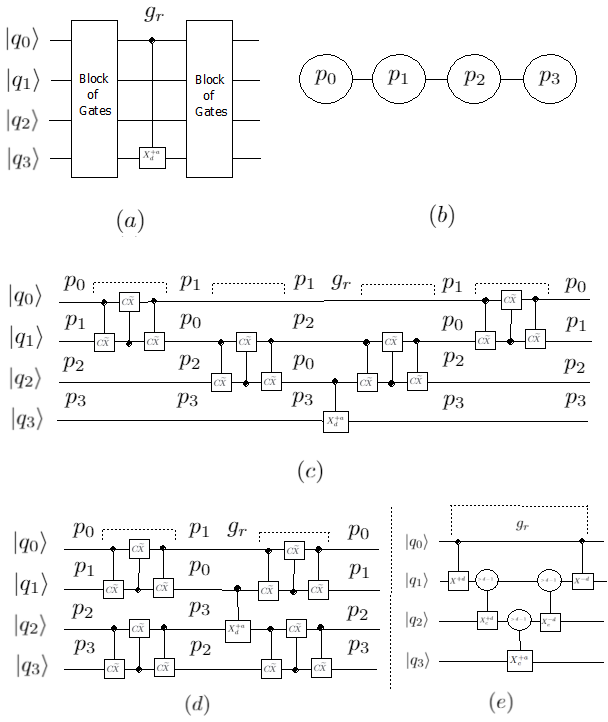}
\caption{$(a)$ Example: circuit; $(b)$ Example: qudit topology; $(c)$ SWAP insertion; $(d)$ SWAP Depth optimization $(e)$ Proposed approach}
\label{exampleswapinsertion}
\end{figure}  

Figure 9(e) shows how the proposed gates can execute the circuit shown in Figure 9(a) based on qudit topology as shown in Figure 9(b) via temporary intermediate higher dimensional qudits. A conventional approach of executing the circuit shown in Figure 9(a) using SWAP for qudit system can be found in Figure 9(c) and Figure 9(d). The initialization of Figure 9(e) can be expressed as 
$\sum_{x,y,z,t = 0}^{d-1}\alpha_{x, y, z, t}\ket{x}\ket{y}\ket{z}\ket{t}$
where the first three are qudits $q_0$, $q_1$ and $q_2$ and the target qudit $t$ is $q_3$,  and $\alpha_{x, y, z, t} \in \mathbb{C}$ and $\sum_{x,y,z,t =0}^{d-1} |\alpha_{x, y, z, t}|^2 = 1$. Here, we show the action of a proposed gates on such a superposition. At first, the $C^{+d}_{X}$ gate operation is performed on the first and the second qudit as illustrated in Equation 27, where first qudit is the control and the second qudit is the target. This upgrades the second qudit to $\ket{d}$ or $\ket{d+1}$ or $\dots$ $\ket{2d-1}$ by availing the higher dimensional space as temporary storage if and only if the first qudit was $\ket{d-1}$. Next, the $C^{+d}_{X_c}$ gate operation is performed on the second and the third qudit as illustrated in Equation 28, where second qudit is the control and the third qudit is the target. This upgrades the third qudit to $\ket{d}$ or $\ket{d+1}$ or $\dots$ $\ket{2d-1}$ by availing the higher dimensional space as temporary storage if and only if the second qudit was $\ket{d}$ or $\ket{d+1}$ or $\dots$ $\ket{2d-1}$. Finally, a  generalized conditional CNOT $C^{+a}_{X_c}$ is applied to the target qudit $q_3$ and the third qudit $q_2$ as control as described in Equation 29. This gate will be executed only when the third qudit were $\ket{d}$ or $\ket{d+1}$ or $\dots$ $\ket{2d-1}$, it would happen only when first qudit was $\ket{d-1}$ state.  The controls are reinstated to their original states by applying $C^{-d}_{X_c}$ gate followed by $C^{-d}_{X}$ gate, which reverses the effect of the first and second gate. Thus the $\ket{d}$ or $\ket{d+1}$ or $\dots$ $\ket{2d-1}$ state from $2d$-ary quantum system can be used instead of SWAP to store temporary information in $q_1$ and $q_2$ qudits, as discussed earlier in binary quantum system as well. As in binary quantum system, for any higher $n$-qudit system, the proposed three gates can be used for moving quantum states for physical implementation without the use of SWAP gate.
\begin{widetext}
\begin{eqnarray}
& & \sum_{x,y,z,t}^{d-1}\alpha_{x, y, z, t}\ket{x}\ket{y}\ket{z}\ket{t}\\
&\stackrel{\mbox{\scriptsize{$C^{+d}_{X}$}}_{q_0,q_1}}{\longrightarrow}& \sum_{x,y,z,t}\alpha_{x\neq d-1,y,z,t}\ket{x \neq d-1}\ket{y}\ket{z}\ket{t} + \sum_{x,y,z,t}\alpha_{x=d-1,y,z,t}\ket{d-1}\ket{(y+d)\%2d}\ket{z}\ket{t}\\
&\stackrel{\mbox{\scriptsize{$C^{+d}_{X_c}$}}_{q_1,q_2}}{\longrightarrow}& \sum_{x,y,z,t}\alpha_{x\neq d-1,y,z,t}\ket{x \neq d-1}\ket{y}\ket{z}\ket{t} + \sum_{x,y,z,t}\alpha_{x=d-1,y,z,t}\ket{d-1}\ket{(y+d)\%2d}\ket{(z+d)\%2d}\ket{t}
\end{eqnarray}
\end{widetext}

\begin{widetext}
\begin{eqnarray}
&\stackrel{\mbox{\scriptsize{$C^{+a}_{X_c}$}}_{q_2,q_3}}{\longrightarrow}& \sum_{x,y,z,t}\alpha_{x \neq d-1,y,z,t}\ket{x}\ket{y}\ket{z}\ket{t} + \sum_{x,y,z,t}\alpha_{x=d-1,y,z,t}\ket{d-1}\ket{(y+d)\%2d}\ket{(z+d)\%2d}\ket{(t+a)\%d}\\
&\stackrel{\mbox{\scriptsize{$C^{-d}_{X_c}$}}_{q_1,q_2}}{\longrightarrow}& \sum_{x,y,z,t}\alpha_{x\neq d-1,y,z,t}\ket{x \neq d-1}\ket{y}\ket{z}\ket{t} + \sum_{x,y,z,t}\alpha_{x=d-1,y,z,t}\ket{d-1}\ket{(y+d)\%2d}\ket{z}\ket{(t+a)\%d}\\
&\stackrel{\mbox{\scriptsize{$C^{-d}_{X}$}}_{q_0,q_1}}{\longrightarrow}& \sum_{x,y,z,t}\alpha_{x\neq d-1,y,z,t}\ket{x \neq d-1}\ket{y}\ket{z}\ket{t} + \sum_{x,y,z,t}\alpha_{x=d-1,y,z,t}\ket{d-1}\ket{(y}\ket{z}\ket{(t+a)\%d}
\end{eqnarray}
\end{widetext}


\begin{table*}[]
\caption{Comparative Analysis}
\centering
\begin{tabular}{ |p{4.9cm}|p{2.6cm}|p{2.6cm}|p{2.6cm}|p{2.6cm}| }
 \hline
 \multirow{2}{4.9cm}{Number of qubits/qudits involved in between two-qubit/qudit gate} & \multicolumn{2}{|c|}{Proposed work} &  \multicolumn{2}{|c|}{Conventional work}\\
 \cline{2-5}
 & Gate count & Depth  & Gate count & Depth \\ \hline
3 & 3 & 3 & 7 & 7\\ \hline
4 & 5 & 5 & 13 & 7\\ \hline
5 & 7 & 7 & 19 & 13\\ \hline
6 & 9 & 9 & 25 & 13\\ \hline
7 & 11 & 11 & 31 & 19\\ \hline
8 & 13 & 13 & 37 & 19\\ \hline
9 & 15 & 15 & 43 & 25\\ \hline
10 & 17 & 17 & 49 & 25\\ \hline
$n$ & $2\times(n-2)+O(1)$ & $2\times(n-2)+O(1)$ & $6 \times (n-2)+O(1)$ & $6 \times (\ceil{\frac{n}{2}}-1) + O(1)$\\ \hline
\end{tabular}
\label{tab:n_controlled}
\end{table*}

\begin{figure*}[ht!]
\centering
\includegraphics[scale=.5]{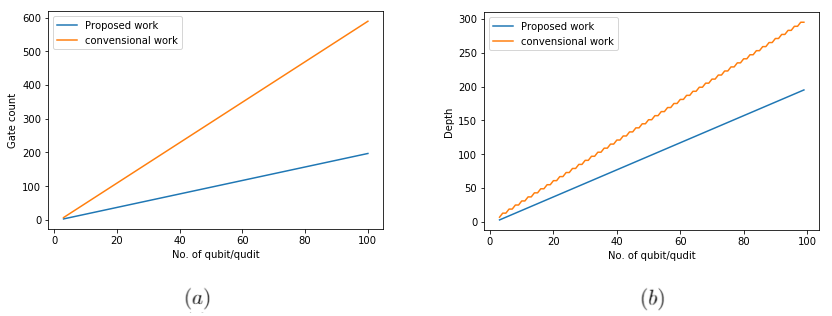}
\caption{$(a)$ Gate count vs No. of qubit/qudit $(b)$ Circuit depth vs No. of qubit/qudit}
\label{gatedepthvsqubit}
\end{figure*}

\section{Discussion and Conclusion}
Here, we tried to implement an universal two-qubit or two-qudit CNOT gate in which qubits/qudits are not adjacent to each other with respect to qubit/qudit topology.  We considered generalized CNOT gate as an example to establish our claim for the simplicity of understanding, even though our approach stands good for any other two-qubit or two-qudit gate with simple modification in the proposed gates as per unitary operation at the target. To make them adjacent, the conventional approach is to insert the SWAP gates between two qubits/qudits. The optimal gate cost of SWAP and its mirror circuit is six CNOT gates using the conventional decomposition-based approach \cite{gidney}, where three qubits are involved as shown in Figure 3. This is also legitimate for any $d$-dimensional quantum system, but the only difference is $C\!\widetilde{X}$ replaces CNOT here. Correspondingly, the optimal depth cost of SWAP and its mirror circuit is six using the conventional decomposition-based approach for any $d$-ary quantum system. Further a four qubits and a four qudits circuit implementation have been illustrated in Figure 5 and Figure 9 respectively. Two SWAP operations have to be inserted to make the circuit executable is shown in Figure 5 and Figure 9. A conventional optimization technique yields the depth constant as compared to the three qubit example as shown in Figure 3. Albeit, gate count increases by six for additional two SWAP insertion. Conventional approach of SWAP insertion generalizes the gate cost and the depth for any $n$-qubit/qudit circuit. For the execution of one two-qubit/qudit gate on $n$-qubit/qudit circuit where $n-2$ SWAP insertions are required to make the control and the target qubit/qudit adjacent, the gate count of the updated circuit becomes $6 \times (n-2)+O(1)$ and the depth of the circuit becomes  $6 \times (\ceil{\frac{n}{2}}-1) + O(1)$ (where $O(1)$ is for two qubit/qudit gate that has to be executed). 

In this paper, we proposed a qubit-qudit approach to move the quantum states through qubits to eradicate the SWAP operation. The higher dimensional quantum states are used as an intermediate states in a qudit system, while the input and output states still remain qubits to solve the nearest neighbour problem. We introduced the $\ket{2}$ and $\ket{3}$ quantum states as temporary storage of quaquad quantum system without hampering the fundamental operation of initialization and measurement on physical devices. Later on the extension of the proposed approach to $d$-dimensional quantum system with the use of $\ket{d}$, $\ket{d+1}$, $\dots$, $\ket{2d-1}$ quantum states of $2d$-ary quantum system as temporary storage has been addressed. For this novel approach, we achieved an optimized gate cost and depth for this problem. As shown in Table 1, we achieved a whooping reduction to $2\times(n-2)+O(1)$ as gate count and $2\times(n-2)+O(1)$ as depth compared to the convention work while $n$ qubits/qudits are involved for a two-qubit/qudit gate. Figure \ref{gatedepthvsqubit} shows that the proposed work outperforms conventional state of the techniques with respect to gate count and circuit depth. 

 The  work has limitations which  can be addressed as described further. Any quantum system is prone to several varied types of errors such as decoherence, noisy gates. For single and two qudit gates the gate error scales as $d^2$ and $d^4$ respectively, for a $d$-dimensional quantum system. Moreover, the amplitude damping error decays the state $\ket{1}$ to $\ket{0}$ with probability $\lambda_1$ for qubits. Every state in level $\ket{i} \neq \ket{0}$ has a probability $\lambda_i$ of decaying  for a $d$-dimensional system. Thus we can say that, the use of higher dimensional states chastises the system with more errors. Gokhale et al. \cite{gokhale2019asymptotic} have shown the effect of these errors on algorithms that uses the temporary intermediate qudits. They have shown that even though the use of qudits  increases error, since the number of gate count and the depth are both reduced the overall error probability of the decomposition is lower than the existing ones. Typical errors are considered by Gokhale et al. \cite{gokhale2019asymptotic} but they did not consider the leakage error. As the $d$-ary system may occasionally need to access states beyond the $d$-ary computational space which is an engineering challenge, it makes the system  susceptible to leakage error \cite{ saha2020asymptotically}. In \cite{ saha2020asymptotically}, the authors have shown the effect of erasure and unitary leakage model for the algorithm that uses intermediate temporary qudits.  Since normal protection schemes against decoherence is unable to correct such a leakage error such leakage is a serious obstacle for reliable computation. 
 
 In future scope of this paper, we would like to mitigate the error that might happen due to the accessibility of higher dimensional space as temporary storage. We would further like to apply our proposed approach to the existing qubit mapping algorithms to demonstrate the usefulness of our novel approach with the benchmarks circuits. In near future, it can also be investigated that the proposed approach may give some advantage in quantum communication as SWAPs are involved there \cite{PhysRevLett.88.127902, PhysRevLett.123.070505, PhysRevLett.125.230501}. The simulation of the proposed circuits for the verification is carried out on Google Colab platform \cite{Bisong2019} and the code is available at \href{https://github.com/amitsaha2806/Moving-Quantum-States-through-Qubits-via-Intermediate-Higher-Dimensional-Qudits}{https://github.com/amitsaha2806/Moving-Quantum-States-through-Qubits-via-Intermediate-Higher-Dimensional-Qudits}.

\nocite{*}


\bibliography{apssamp}



\end{document}